\documentclass[preprint,3p,times]{elsarticle}

\usepackage{amssymb}
\usepackage{amsmath}

\usepackage{lineno} 

\usepackage{orcidlink} 
\usepackage{comment} 
\usepackage{gensymb} 

\journal{Nucl. Instrum. Methods Phys. Res. A} 

\begin{document}

\begin{frontmatter}



\title{Ultra-pure Nickel for Structural Components of Low-Radioactivity Instruments} 


%
%


\author[PNNL]{T.~J.~Roosendaal}
\author[PNNL]{C.~T.~Overman}
\author[PNNL]{G.~S.~Ortega\textsuperscript{\orcidlink{0000-0002-4685-1826}}}
\author[PNNL]{T.~D.~Schlieder}
\author[PNNL]{N.~D.~Rocco\textsuperscript{\orcidlink{0000-0001-9420-073X}}}
\author[PNNL]{L.~K.~S.~Horkley}
\author[PNNL]{K.~P.~Hobbs\textsuperscript{\orcidlink{0009-0004-8537-2811}}}
\author[PNNL]{K.~Harouaka\textsuperscript{\orcidlink{0000-0002-5305-0138}}}
\author[PNNL]{J.~L.~Orrell\textsuperscript{\orcidlink{0000-0001-7968-4051}}\corref{cor1}}\ead{john.orrell@pnnl.gov} 

\cortext[cor1]{Corresponding author}

\author[Alabama]{P.~Acharya}
% ...


\affiliation[PNNL]{{Pacific Northwest National Laboratory, Richland, WA 99352, USA}}
\affiliation[Alabama]{{Department of Physics and Astronomy, University of Alabama, Tuscaloosa, AL 35405, USA}}
% ...


\begin{abstract}

The next generation of rare‐event search experiments in nuclear and particle physics demand structural materials combining exceptional mechanical strength with ultra‐low levels of radioactive contamination. This study evaluates chemical vapor deposition (CVD) nickel as a candidate structural material for such applications.
Manufacturer-supplied CVD Ni grown on aluminum substrates underwent tensile testing before and after welding alongside standard Ni samples. CVD Ni exhibited a planar tensile strength of $\sim$600~MPa, significantly surpassing standard nickel. However, welding and heat treatment were found to reduce the tensile strength to levels comparable to standard Ni, with observed porosity in the welds likely contributing to this reduction. 
Material assay via inductively coupled plasma mass spectrometry (ICP-MS) employing isotope‐dilution produced measured bulk concentration of $^{232}$Th, $^{238}$U, and $^{\text{nat}}$K at the levels of $\sim$70~ppq, $\lesssim$100~ppq, and $\sim$900~ppt, respectively, which is the lowest reported in nickel. Surface‐etch profiling uncovered higher concentrations of these contaminants extending $\sim$10~$\mu$m beneath the surface, likely associated with the aluminum growth substrate. 
The results reported are compared to the one other well documented usage of CVD Ni in a low radioactive background physics research experiment and a discussion is provided on how the currently reported results may arise from changes in CVD fabrication or testing process. These results establish CVD Ni as a promising low-radioactivity structural material, while outlining the need for further development in welding and surface cleaning techniques to fully realize its potential in large-scale, low radioactive background rare‐event search experiments.

\end{abstract}




\end{frontmatter}



\begin{twocolumn} 

\section{Introduction}\label{sec:Intro}

nEXO is a concept for a tonne-scale neutrinoless double beta decay experiment consisting of 5~tonnes of liquid xenon, enriched to 90\% $^{136}$Xe, and operated as a time projection chamber for detection of radiation-induced ionization events~\cite{nEXO_Adhikari_2022}. Observation of neutrinoless double beta decay would imply neutrinos are Majorana fermions~\cite{Majorana2006}, which would be a unique characteristic among the particles within the Standard Model. This observation would also demonstrate the existence of lepton number violation and have potential ramifications for understanding the mat\-ter/anti\-mat\-ter a\-sym\-metry of the universe~\cite{WOS:000492825700009}. A critical design characteristic for nEXO is the use of construction materials that are extremely low in radioactive contaminants which may generate backgrounds to the observation of the rare double beta decay events of $^{136}$Xe~\cite{nEXO-PreCDR}. In this report, the investigation of ultra-pure chemical vapor deposition (CVD) nickel (Ni) is presented. In particular, the nEXO experiment initially pursued use of CVD Ni for the construction of the inner and outer vessels of the cryostat used to maintain the detector volume at liquid xenon temperatures. The spherical inner and outer vessels have nominal diameters of 3.4~m and 4.5~m, respectively. The purpose and size of these components implies the need for both strength and low levels of radioactive contaminants of the CVD Ni.

To the authors' best knowledge, the first (and last to date\footnote{The SNO NCDs were repurposed for the HALO experiment~\cite{HALO_2008}.}) use of ultra-pure CVD Ni for low-radioactive background rare-event searches in nuclear and particle physics, was in the construction of the Sudbury Neutrino Observatory's (SNO's)~\cite{BOGER2000172,PhysRevC.88.025501} Neutral Current Detectors (NCDs)~\cite{SNO-NCD-IEEE-1999,SNO-NCD-NIMA-2007}. The SNO NCDs were cylindrical $^3$He proportional counters with 5.08~cm outer diameter and lengths ranging from 200--300~cm. A key factor in the choice of CVD Ni was the low concentrations of radioactive $^{232}$Th- and $^{238}$U-decay chain nuclides which could contribute to a background event rate in the SNO experiment. However, as noted by the SNO Collaboration~\cite{SNO-NCD-IEEE-1999}, the CVD Ni tubes were produced on Al mandrels, and that Al (as well as trace materials within the Al) can be incorporated into the surface of the CVD Ni. Thus, the SNO Collaboration used post-production electropolishing and etching as a surface cleaning step. The CVD Ni was produced by the Canadian company Mirotech, Inc. and later Chemical Vapour Deposition Systems, Inc., which acquired Mirotech~\cite{SNO-NCD-NIMA-2007}. Further details of the physical properties and radioactivity concentrations of the SNO NCD CVD Ni are referenced later in this report for comparison purposes.

This report first briefly describes the production of samples from the CVD Ni obtained by the nEXO Collaboration for analysis (Sec.~\ref{Sec:Samples}). The contents of this report focus on key measured physical properties (Sec.~\ref{Sec:Strength}) and trace rad\-io\-ac\-tive con\-tam\-i\-na\-tion levels (Sec.~\ref{Sec:Assay}). The physical property analyses focus on tensile strength before and after both welding and heat treatment, where comparison is provided against commercial off the shelf (COTS) Ni samples. The trace radioactive contamination analysis focuses on determining the bulk and surface concentrations of $^{232}$Th, $^{238}$U, and, for the first time reported, $^{40}$K. To place this report's results in context, at the end of both the physical properties and trace radioactive contamination sections (Sec.~\ref{Sec:Strength} and~Sec.~\ref{Sec:Assay}, respectively), comparisons to the information available from the SNO Collaboration's experience with CVD Ni are presented. A brief conclusion (Sec.~\ref{Sec:Conclusion}) summarizes the results and recommendations for further development of CVD Ni as a generic ultra-pure construction material for use in future nuclear and particle physics rare-event searches requiring low-radioactive background environments.

\section{CVD Ni Samples}\label{Sec:Samples}

The CVD Ni samples used in the studies described in this report were fabricated by CVMR (Tor\-on\-to, Can\-a\-da)~\cite{CVMR}, formerly known as Chemical Vapour Deposition Systems, Inc. The two CVD Ni panels produced by CVMR were fabricated via a Mond CVD process~\cite{MondCVD-1890} on Al substrates at approximately 200~$\degree$C. Pri\-or re\-ports~\cite{SNO-NCD-IEEE-1999} state the CVD Ni grows on the substrate at 0.75~mm/hr. Thicknesses on the scale of 5~cm are readily achieved. After formation, the higher coefficient of thermal expansion of aluminum compared to nickel causes the nickel to separate from the aluminum substrate when cooled. The samples used in this study were received at PNNL in October 2023.

The CVD Ni panels produced by CVMR were nominally 155~mm square, with a measured thickness that varied from 6.3--7.3~mm. The two panels were connected by a thin nickel growth and outer rim as seen in Figure~\ref{fig:as_received}. The growth surface of the nickel was smooth and contained a small number of low-profile spherical protrusions, while the bottom of the sample attached to the substrate matched the finish of the Al substrate. In preparation for sampling, the two panels were separated and the thicknesses were machined down to nominally 5.4~mm.

\begin{figure}
    \centering
    \includegraphics[width=0.75\linewidth]{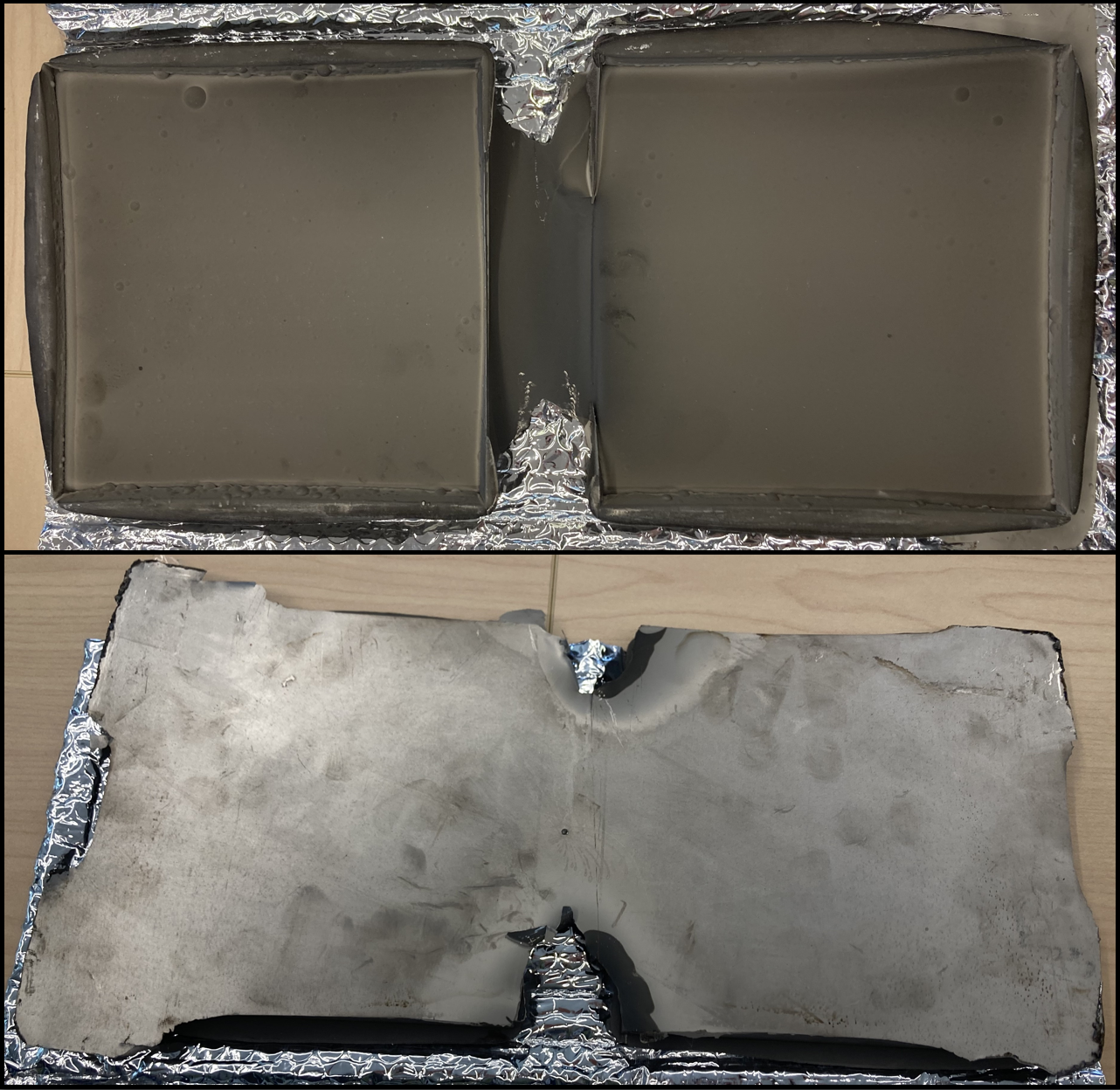}
    \caption{As received CVMR Nickel panels. Top: Growth surface, Bottom: Substrate surface. The substrate was Al (not shown).}
    \label{fig:as_received}
\end{figure}

The two panels were each cut into 2 weld sections, 3 weld filler rods, and 1 assay transect section as shown in Figure~\ref{fig:sampling_layout}. The left and right panels in Figure~\ref{fig:as_received} were sampled in the vertical and horizontal directions, respectively, to ensure assay samples were taken from a large spatial area. Three weld sections were divided in half again, along the dashed lines in Figure~\ref{fig:sampling_layout}, and prepared with full thickness 40~degree chamfer weld prep. The fourth section was retained ``as supplied'' to provide un-welded control samples for tensile testing. The two assay transect sections were sub-sampled into sixteen 1~g and ten 5~g samples. The sample counts described above are shown within the process diagram (Fig.~\ref{fig:flow_diagram}) described below.

\begin{figure}
    \centering
    \includegraphics[width=0.95\linewidth]{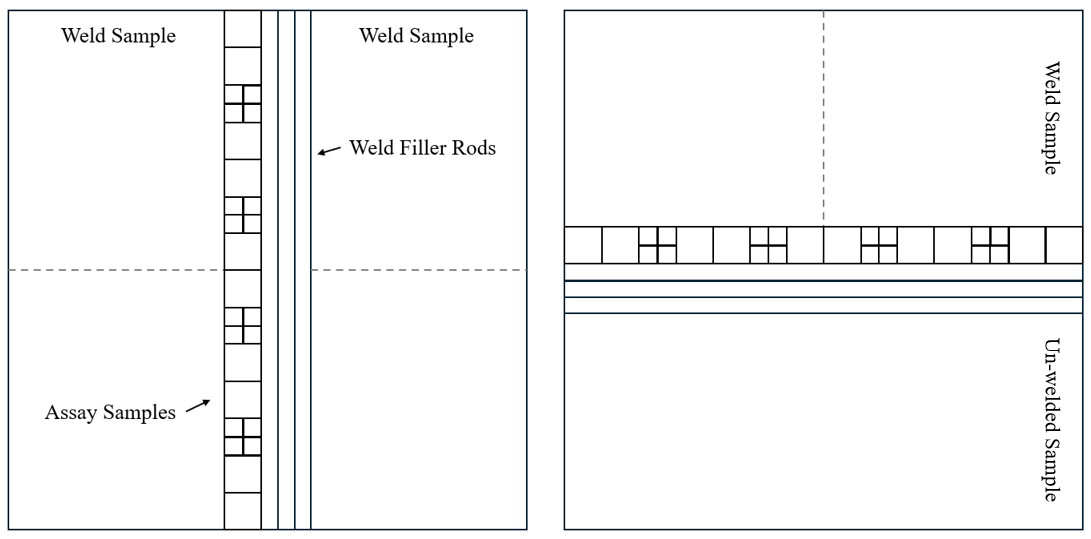}
    \caption{Vertical (left) and horizontal (right) sampling patterns for the two CVD Ni panels.}
    \label{fig:sampling_layout}
\end{figure}

The left-to-right flow diagram in Figure~\ref{fig:flow_diagram} shows the process used to assess the CVD Ni and generate the results presented in this report. Rough machining leads to the sample sectioning described above (see Fig.~\ref{fig:sampling_layout}). Along the top of the flow diagram, the single un-welded CVD Ni section is used to create two dog-bone samples as an ``as supplied'' control sample. The three weld samples are used to create six dog-bone samples. Together these eight dog-bone samples are tested for tensile strength and then x-ray interrogated (see below for x-ray details). The subsize E8 dog-bone samples were cut from the grips of the larger dog-bone samples, see Figures~\ref{fig:DogBones} and~\ref{fig:DogBones_small} for further detail. The subsize E8 dog-bone samples were used in the heat treatment study (a\-long with non-heated control samples) as described below. Along the bottom of the flow diagram, the samples used for trace-level radioactive contamination are shown. Great\-er detail on the processing of the assay samples is provided in Section~\ref{Sec:Assay} on material assay.

\begin{figure}
    \centering
    \includegraphics[width=1\linewidth]{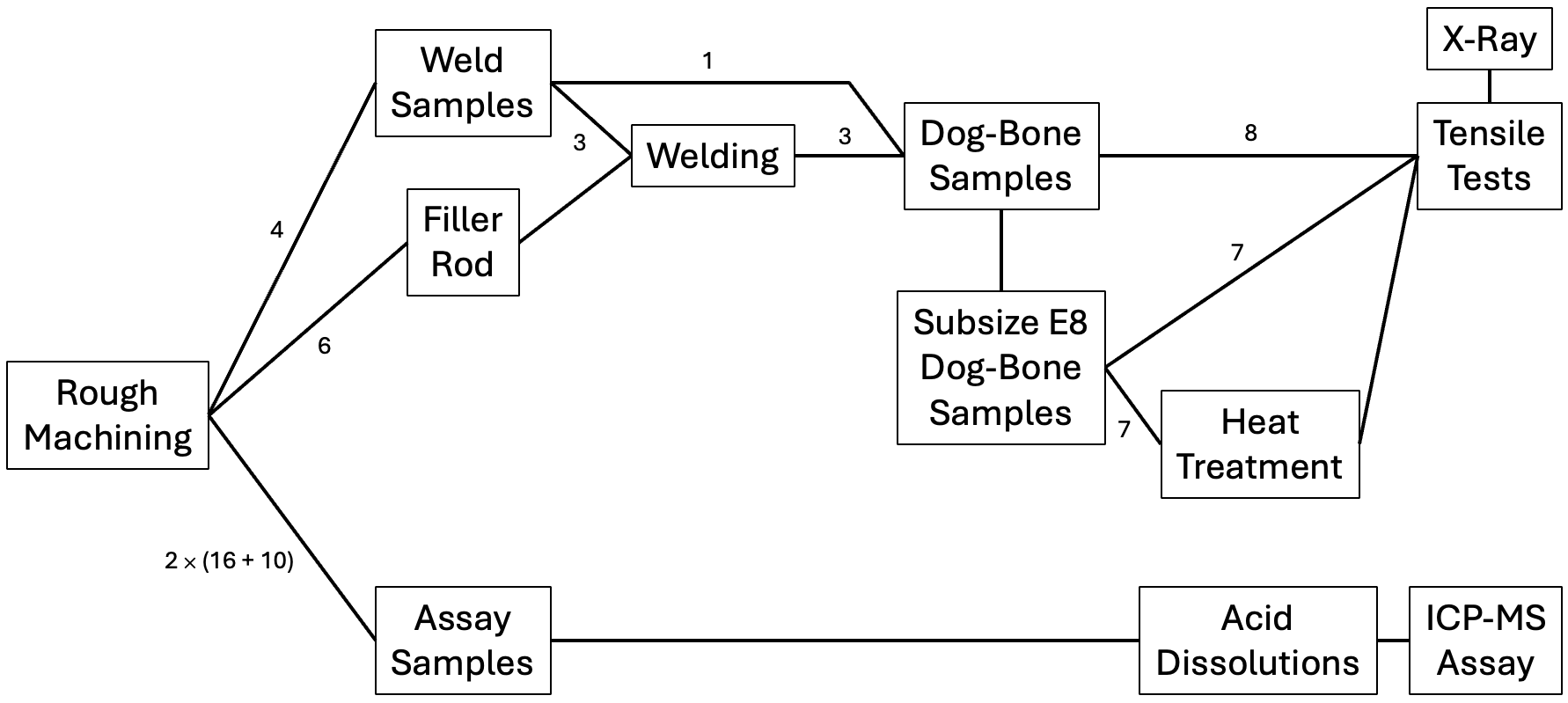}
    \caption{Diagram showing processing steps. The process flows from left to right. Numbers alongside the lines are the sample count at each stage. See Sec.~\ref{Sec:Samples} for a description of the process.}
    \label{fig:flow_diagram}
\end{figure}

To provide comparisons to COTS Ni, plates of Ni-200 were procured from OnlineMetals.com, part number 10636, with dimensions 0.25~in $\times$ 12~in $\times$ 12~in. This COTS Ni was specified as Ni sheet/plate conforming to ASTM-B162 material specifications. The COTS Ni, unless otherwise noted, has been treated the same as the CVD Ni for all strength-evaluation processing steps shown in Figure~\ref{fig:flow_diagram}.


\section{Tensile Strength}\label{Sec:Strength}

Low-radioactivity construction materials must have appropriate strength, and demonstrate uniformity in response to assembly processing, if they are to be used in the structures of large-scale instruments. It is anticipated that welding will be the preferred joining technique for CVD Ni components. Tungsten inert gas (TIG) welding was selected for this study because TIG would permit more versatile ``fabrication in place'', though other methods (\textit{e.g.,} electron beam welding) could be considered, depending upon the application. To assess the attributes of CVD Ni for structural components, tensile strength was investigated for the CVD Ni both pre- and post-processing, as described below.

\subsection{Methods}\label{Sec:Methods}

Tensile-tests provide the primary metric for strength evaluation of the welding and heat-treatment processing methods considered in this study. The standardized methods used for these investigations are described in the following paragraphs.

To create the un-welded and post-welded dog-bone samples for the tension pull tests, samples were prepared with the following dimensions: 32~mm grip width, 152~mm total length, 20~mm wide and 33~mm long gauge section. These dimensions were chosen based on ASME welding, brazing, and fusing qualifications standard BPVC.IX-2023, Figure QW-462.1(a). See Figure ~\ref{fig:DogBones} for a photograph of seven\footnote{Eight samples were created. Sample B1 became a ``showpiece'' and was not available for photography or the heat treatment study.} of the CVD Ni dog-bone samples after the pull-tests. There were two dog-bones created from each weld for a total of six welded dog-bone samples of CVD Ni (sample notation: A1, A2, B1, B2, C1, and C2) as well as two dog bone samples from un-welded CVD Ni (sample notation: D1 and D2). A second set of tensile test samples was prepared to test the effects of heat treating the CVD Ni, see Figure ~\ref{fig:DogBones_small}. The subsize E8 samples were prepared with the following dimensions: 4.76~mm grip width, 38.10~mm total length, 3.18~mm wide and 12.70~mm long gauge section. The sample notation follows the same naming convention above with heat treated samples appended with ``H''. In all cases the dog bones were created parallel to the plane of the forming substrate, resulting in all measurements producing planar ($\parallel$) measurement values. See~\cite{Bansa2001} for details on CVD Ni properties relative to planar ($\parallel$) versus transverse ($\perp$) growth directions.

\begin{figure}
    \centering
    \includegraphics[width=1\linewidth]{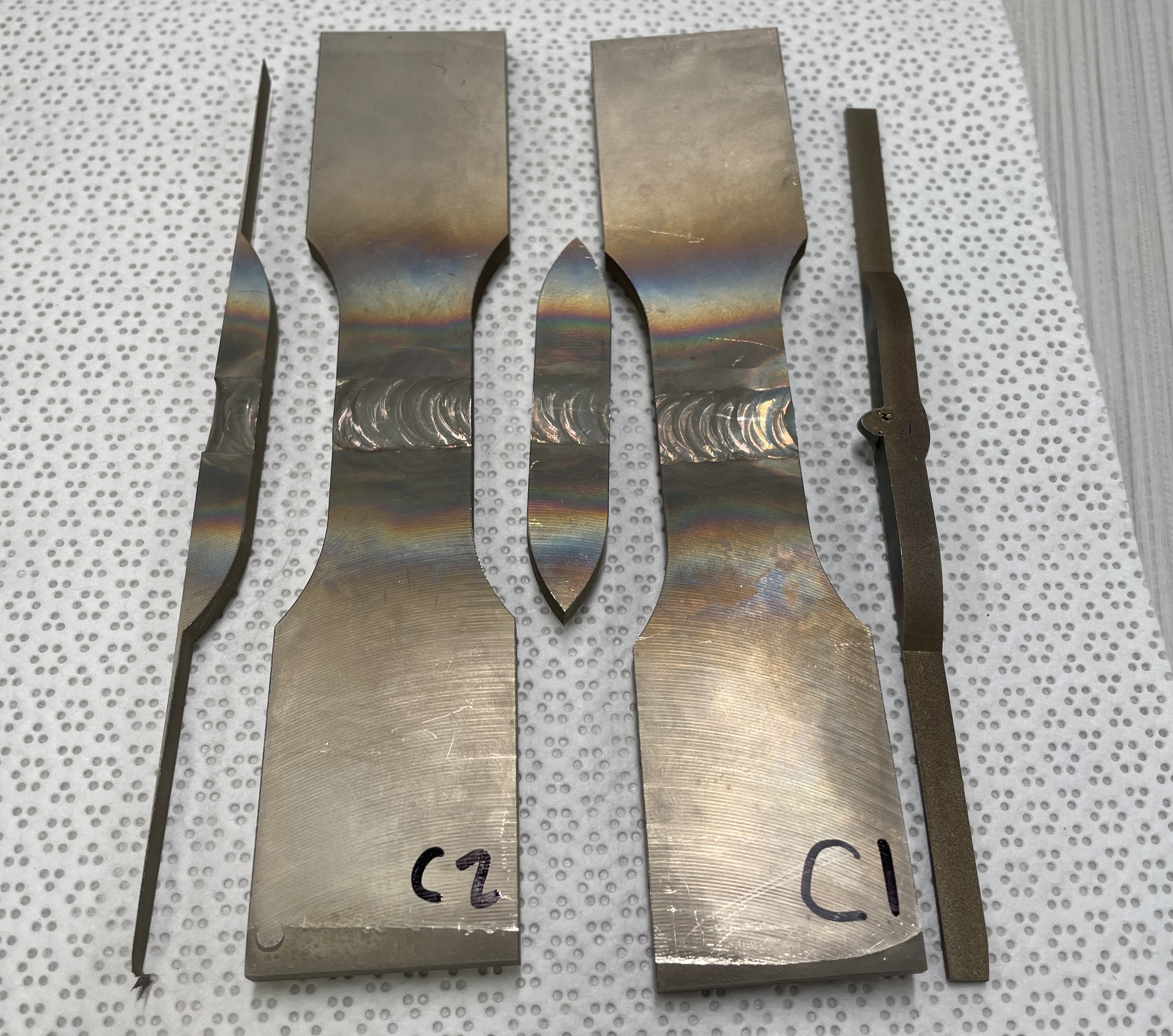}
    \includegraphics[width=1\linewidth]{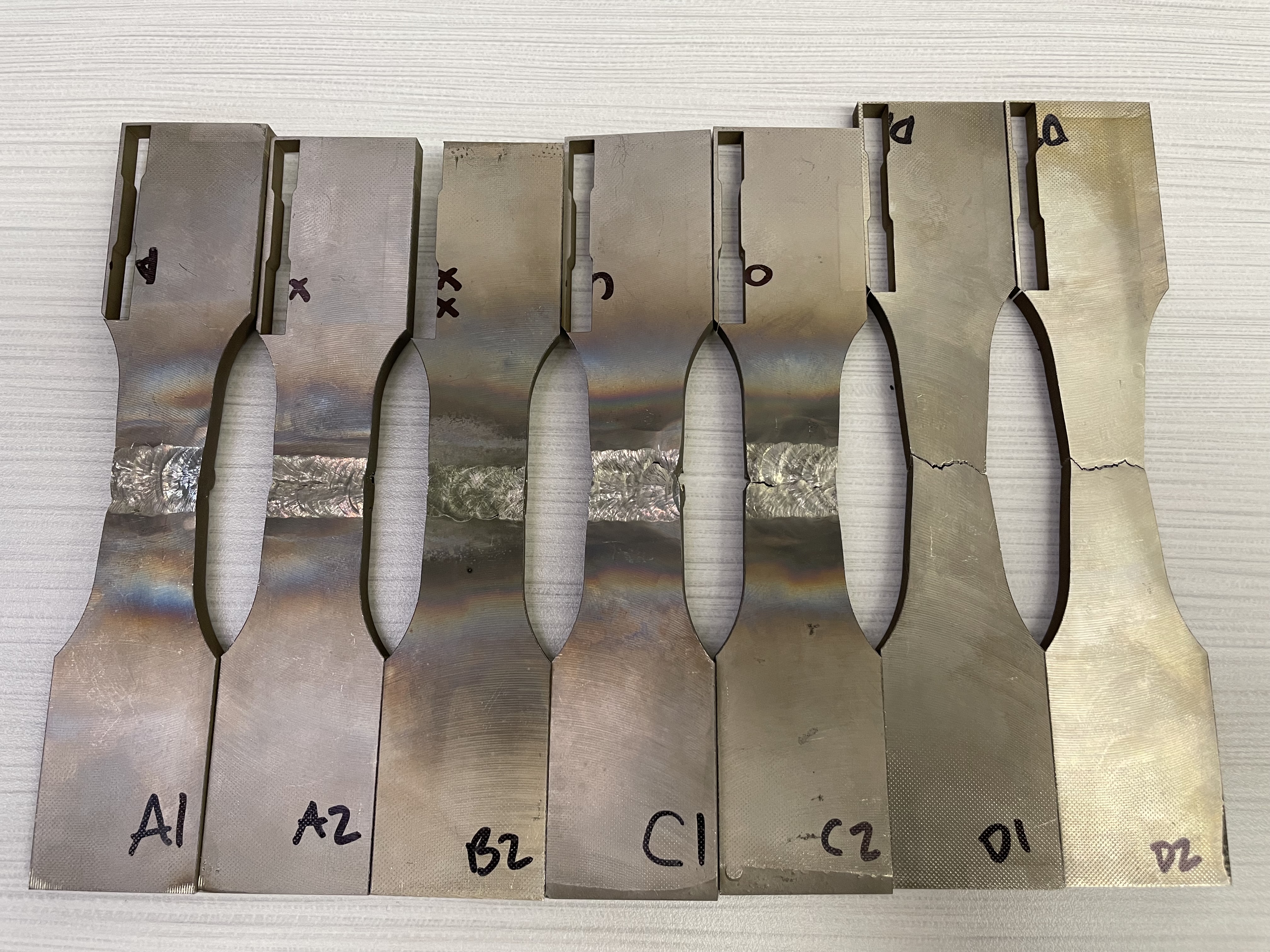}
    \caption{Photographs of the dog-bone sampling after welding (upper panel) and of seven CVD Ni dog-bone samples after tensile testing (lower panel). The subsize E8 samples (see Fig.~\ref{fig:DogBones_small}) were cut from the grip section of these samples, as visible in the upper left of each.}
    \label{fig:DogBones}
\end{figure}

\begin{figure}
    \centering
    \includegraphics[width=1.0\linewidth]{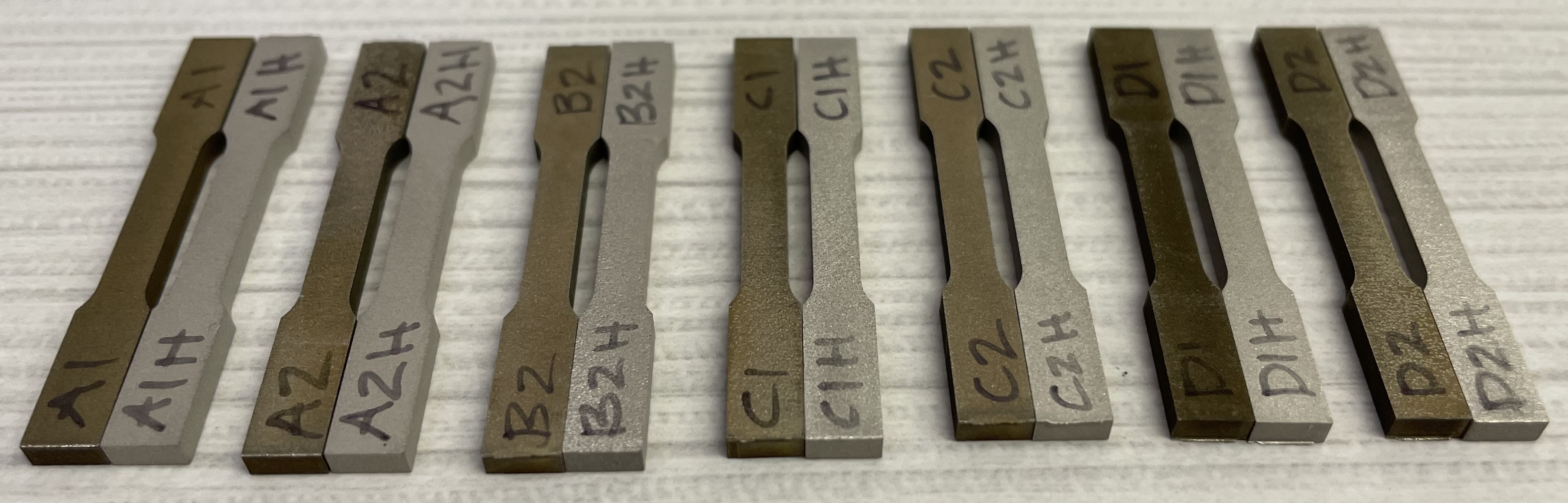}
    \caption{Photograph of seven pairs of CVD Ni prepared in subsize E8 dog-bone tensile test samples. For each pair the non-heat-treated (left) and heat-treated (right) sample is shown. These subsize dog-bone samples were taken from the full-size dog-bones (as seen in Fig.~\ref{fig:DogBones}), and cut in the plane, resulting in the subsize E8 dog-bones being roughly half as thick as the full size dog-bone samples.}
    \label{fig:DogBones_small}
\end{figure}

Tensile testing was performed on an MTS 312.21 frame controlled by Instron Bluehill2 software. A 5,000~lb capacity load cell was used to measure force on the specimen while it was clamped by hydraulic wedge grips. Strain was measured with an Epsilon ONE-78PT-200 extensometer across marks placed near the ends of the gauge section.  Each specimen was tested at a constant displacement rate of 0.025~in/min (0.635~mm/min) to failure. Data was collected at 10~Hz. Tensile strength, yield, strain at break, and modulus were calculated automatically within the Bluehill2 software.

The welding set-up included a controlled Ar environment used during the pulsed TIG welding. Welding was performed using a commercial pure tungsten tip and with 1/8-in (3.175~mm) wide square cross-section fill rods made of the same CVD or COTS Ni (matched to the respective materials to be joined). The voltage was set between 17-19~V with the current maximum set to 200~A. The average current draw while welding was 140~A. There was increased difficulty in obtaining a clean root pass on which to start the weld with the CVD Ni, compared to COTS Ni. After the root pass the remainder of the fill passes went well. Samples were beveled at 40-deg, and to simulate welding nEXO vessel sections, all welding was required to be performed from the top side of the material. This means that a backing weld that would normally be used for material of this thickness could not be used. Furthermore, no grinding or machining of the weld surfaces was allowed after the parts had been acid-etched to represent the anticipated cleanliness protocols required for nEXO or similar experiments. Due to this restriction, the welds were not made flush with the base metal through grinding or machining as is standard for tensile test specimens.

As described in the results section below, the welding efforts raised concerns about the impact of heat (annealing) on the strength of the CVD Ni. To address this, an oven heat-treatment evaluation process was performed. The subsize E8 CVD Ni dog-bone samples were prepared as described above and subjected to heat treatment in an Ar atmosphere using a Camco furnace that was ramped from room temperature to 1000~$\degree$C over a 15~minute period, held at 1000~C for 4~minutes, and then allowed to cool naturally back to room temperature in the furnace over a 1~hour period. This was intended to simulate heating and cooling during welding without melting; generating a void free sample with the TIG weld strength that one might expect to achieve with CVD Ni under natural cooling conditions.

To provide comparison, COTS Ni was processed through the same steps described above for the CVD Ni. This provides COTS Ni comparisons for the welding process, and tensile strength in both un-welded and welded samples. Thus, COTS Ni plays a substantive role for comparing the results presented in the follow sections on strength evaluation. The dog-bone sample notation for COTS Ni was E1, E2, F1, and F2 for welded samples and G1 and G2 for the un-welded samples. COTS Ni was not included in the subsize E8 heat-treat evaluations.

\subsection{Strength Evaluation}\label{Sec:StrengthResults}

Figure ~\ref{fig:RawPullCurves} shows the tensile test curves for samples from the manufacturer-supplied, un-welded CVD Ni and COTS Ni. The two un-welded CVD Ni samples experienced higher tensile stress and broke at roughly 40\% of the displacement distance compared to the two COTS Ni samples. Figure ~\ref{fig:RawPullBreaks} shows the visually distinctive break-formation for these un-welded CVD Ni and COTS Ni samples. In particular, the CVD Ni shows almost no necking at the break compared to the COTS Ni, which is consistent with the differing displacements shown in the tensile test curves (Fig.~\ref{fig:RawPullCurves}). From the strain and separation curves, the tensile strength is determined and presented in summary Table~\ref{tab:AllPullData}. These results of $\sim$600~MPa planar ultimate tensile for pure CVD Ni are consistent with the results from a separate, extensive physical properties evaluations of CVD Ni~\cite{Bansa2001}, (see their Table~4.6, pg. 75).

\begin{figure}
    \centering
    \includegraphics[width=1\linewidth]{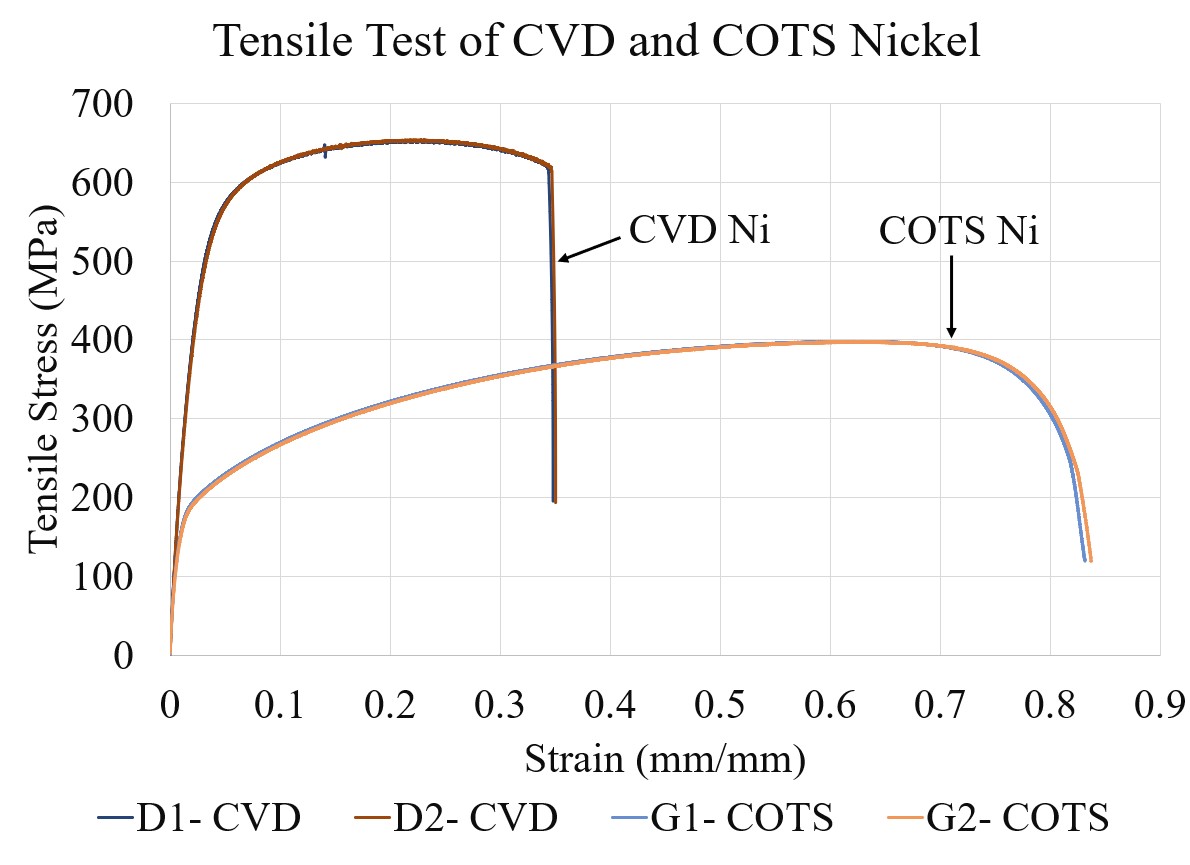}
    \caption{Tensile test curves for un-welded CVD and COTS Ni samples. The curves for the two CVD Ni samples are nearly identical. The curves for the two COTS Ni only deviate from one another at the highest strain values.}
    \label{fig:RawPullCurves}
\end{figure}

\begin{figure}
    \centering
    \includegraphics[width=1\linewidth]{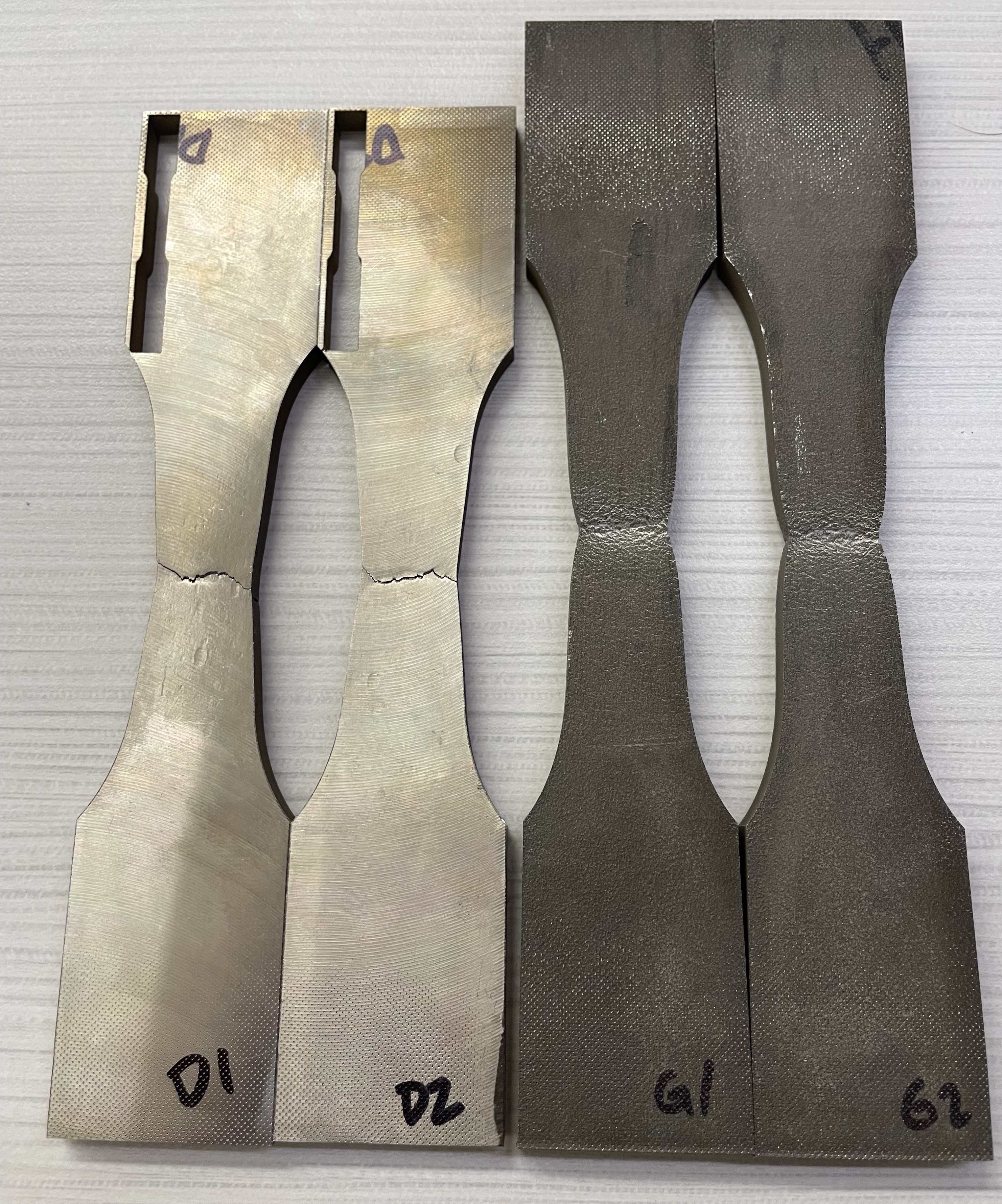}
    \caption{Un-welded COTS (G2 and G1) and CVD (D2 and D1) Ni samples after tensile testing. Notice the difference in break pattern between the materials.}
    \label{fig:RawPullBreaks}
\end{figure}

Figure ~\ref{fig:WeldedPullCurves} shows the tensile test curves for samples prepared from the welded CVD Ni and COTS Ni. From the strain and separation curves, the tensile strength is determined and presented in summary Table~\ref{tab:AllPullData}.

\begin{figure}
    \centering
    \includegraphics[width=1\linewidth]{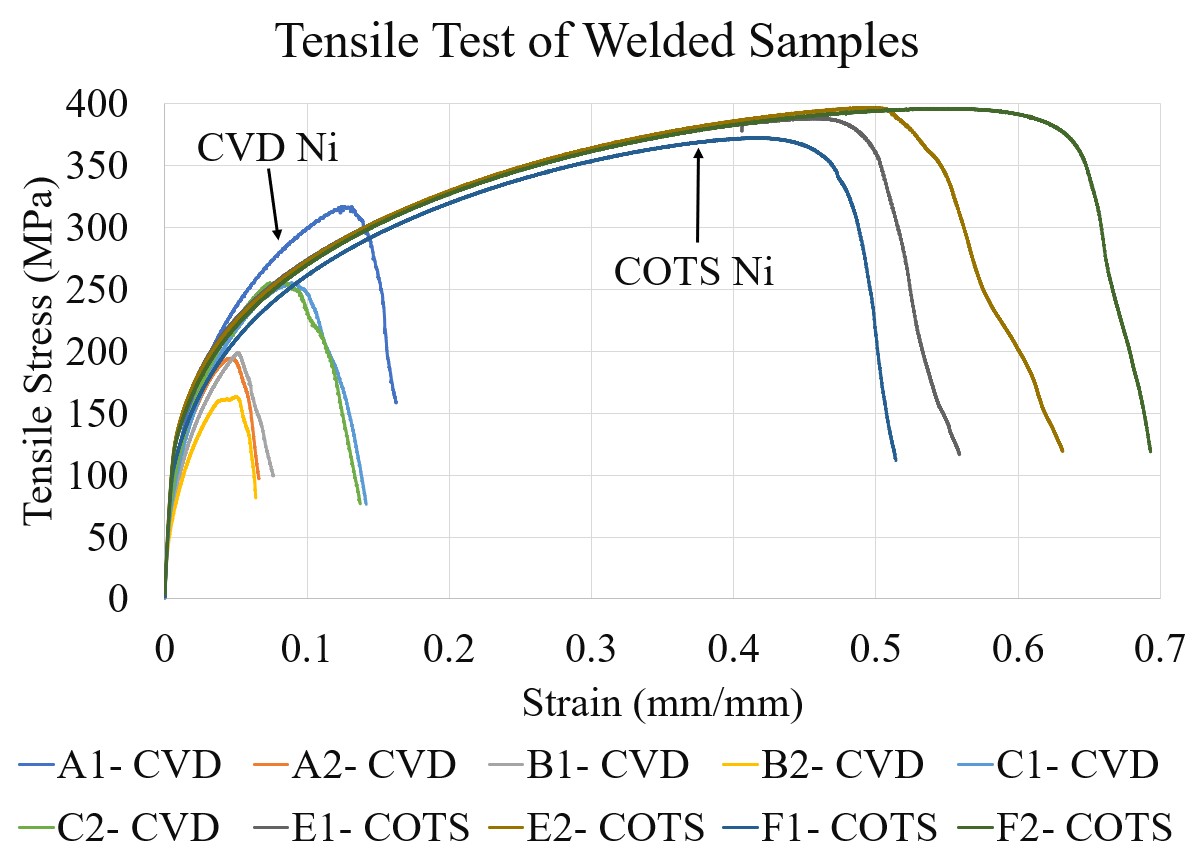}
    \caption{Tensile test curves for welded CVD and COTS Ni samples. CVD Ni shows reduced tensile strength after welding; compare to Fig.~\ref{fig:RawPullCurves}.}
    \label{fig:WeldedPullCurves}
\end{figure}

\renewcommand{\arraystretch}{1.2}
\begin{table}
\centering
\begin{tabular}{| c | c | c | c |} 
\hline 
Sample & Maximum & Ultimate & Displacement \\
        & Force   & Tensile  & at Break     \\
        &         & Strength & (standard)   \\
        & [kN]    & [MPa]    & [mm]         \\
\hline
\hline
\multicolumn{4}{|c|}{CVD Ni} \\ \hline
A1 & 34.15 & 316.45 & 5.37 \\ 
A2 & 20.90 & 193.98 & 2.18 \\
B1 & 21.50 & 198.52 & 2.52 \\
B2 & 17.69 & 163.16 & 2.10 \\
C1 & 27.33 & 254.80 & 4.67 \\
C2 & 27.60 & 256.12 & 4.53 \\ \hline
D1 & 70.40 & 652.73 & 11.36 \\
D2 & 70.66 & 654.18 & 11.45 \\
\hline
\hline
\multicolumn{4}{|c|}{COTS Ni} \\ \hline
E1 & 51.72 & 388.08 & 18.44 \\
E2 & 52.99 & 396.64 & 20.83 \\
F1 & 49.43 & 372.22 & 16.96 \\
F2 & 52.79 & 395.90 & 22.87 \\ \hline
G1 & 53.19 & 397.96 & 27.44 \\
G2 & 53.18 & 397.48 & 27.63 \\
\hline
\end{tabular}
\caption{Numerical data from the tensile strength analysis. The displacement at break is measured crosshead. Samples A--C and E--F are for welded assemblies, as described in Section~\ref{Sec:Methods}. The D and G samples are ``as supplied'' (un-welded).}
\label{tab:AllPullData}
\end{table}
\renewcommand{\arraystretch}{1.0}

Comparison of the pull test curves (Fig.~\ref{fig:RawPullCurves} and~\ref{fig:WeldedPullCurves}) as well as the summarized results data (Tab.~\ref{tab:AllPullData}) show several consistent, comparative features. In all cases the COTS Ni---\emph{both welded and un-welded}---shows a maximum tensile stress of approximately 400~MPa. The CVD Ni always breaks after a substantially shorter displacement. Perhaps most interesting for future design considerations is the significant reduction (a factor of 2 or greater) in maximum tensile strain of the welded CVD Ni compared to the un-welded CVD Ni (see Fig.~\ref{fig:RawPullCurves} and~\ref{fig:WeldedPullCurves}). Also noteworthy is that the yield strength is poorly defined for the welded samples. All together, these results and observations of the welded samples warrant the heat treatment study described below.

\subsection{Welding Observations}\label{Sec:WeldingObs}

After welding the CVD Ni plates the tensile samples were cut using an electrical discharge machining (EDM) unit. These cuts through the welded section revealed large void spaces up to 1~mm in diameter. X-ray radiography was used to scan the welds. The x-ray system configuration included use of a 160~kVp Brems\-strah\-lung beam, 2~mm Cu filtration, and 5~mA beam current. First scans used no filtration, but contrast was low. Detector total exposure time was 20~s, averaged over ten, 2~s exposures.

Figure~\ref{fig:radiography} presents imaging scans showing widespread voids through one sample (B) and focused sections of voids in the other samples (A and C) along the root pass of the weld. These voids were not observed in the COTS Ni samples, except for a thin void corresponding to a point where a weld was stopped and then finished on a later date.

\begin{figure}
    \centering
    \includegraphics[width=1\linewidth]{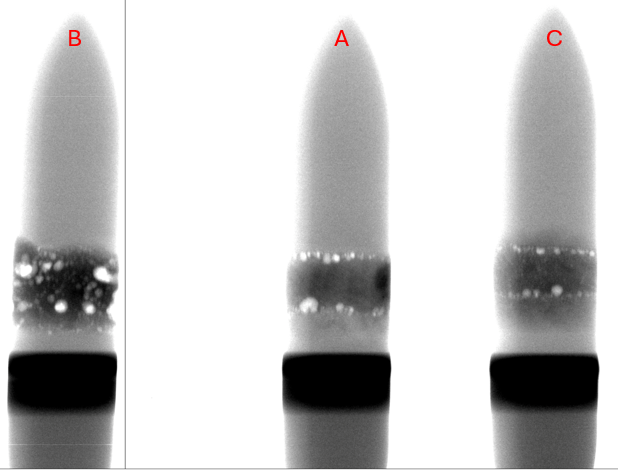}
    \caption{X-ray radiograph of CVD Ni weld sections showing the presence of voids that were not observed in the COTS Ni weld samples. The x-ray analysis was performed on portions of the weld section originating from between the dog-bone samples, see the upper panel of Fig.~\ref{fig:DogBones}. See Sec.~\ref{Sec:WeldingObs} for further details on the x-ray analysis evaluation.}
    \label{fig:radiography}
\end{figure}

It is noted that the CVD Ni samples underwent an acid etch cleaning process similar to what would be done during the nEXO assembly process. This acid etch may have oxidized the surface of the samples, as was identified during the assay process. The COTS Ni samples did not undergo the acid etch process, and the welder was allowed to grind the weld surfaces directly prior to welding. These points are worthy of consideration in planning future CVD Ni welding development.

\subsection{Heat Treatment Effect}\label{Sec:HeatTreat}

The observations of porosity in the CVD Ni welds as well as the decrease in CVD Ni tensile strength after welding was seen as a potential heat-related effect. To explore this possibility, the subsize E8 samples were created (Sec.~\ref{Sec:Samples}) for heat treatment testing (Sec.~\ref{Sec:Methods}), as described above.

Samples that were heat treated changed color from a darker color to a brighter color, similar to the color after being cleaned and acid etched, as is readily seen in Figure~\ref{fig:DogBones_small}.

Tensile tests showed a significant difference in performance between the heat treated and non-treated samples. See Figure~\ref{fig:small_pull} for the pull test curves for these subsize E8 CVD Ni samples. Heat treated samples showed a reduced ultimate tensile stress, as well as much more uniform performance across samples. It is notable that the ultimate tensile stress of the heat treated CVD Ni is on average around 400~MPa, which is quite similar to the ultimate tensile stress shown by the un-welded COTS Ni. Non-treated CVD Ni samples showed a higher ultimate tensile stress, but also showed more variability in those results.

\begin{figure}
    \centering
    \includegraphics[width=1\linewidth]{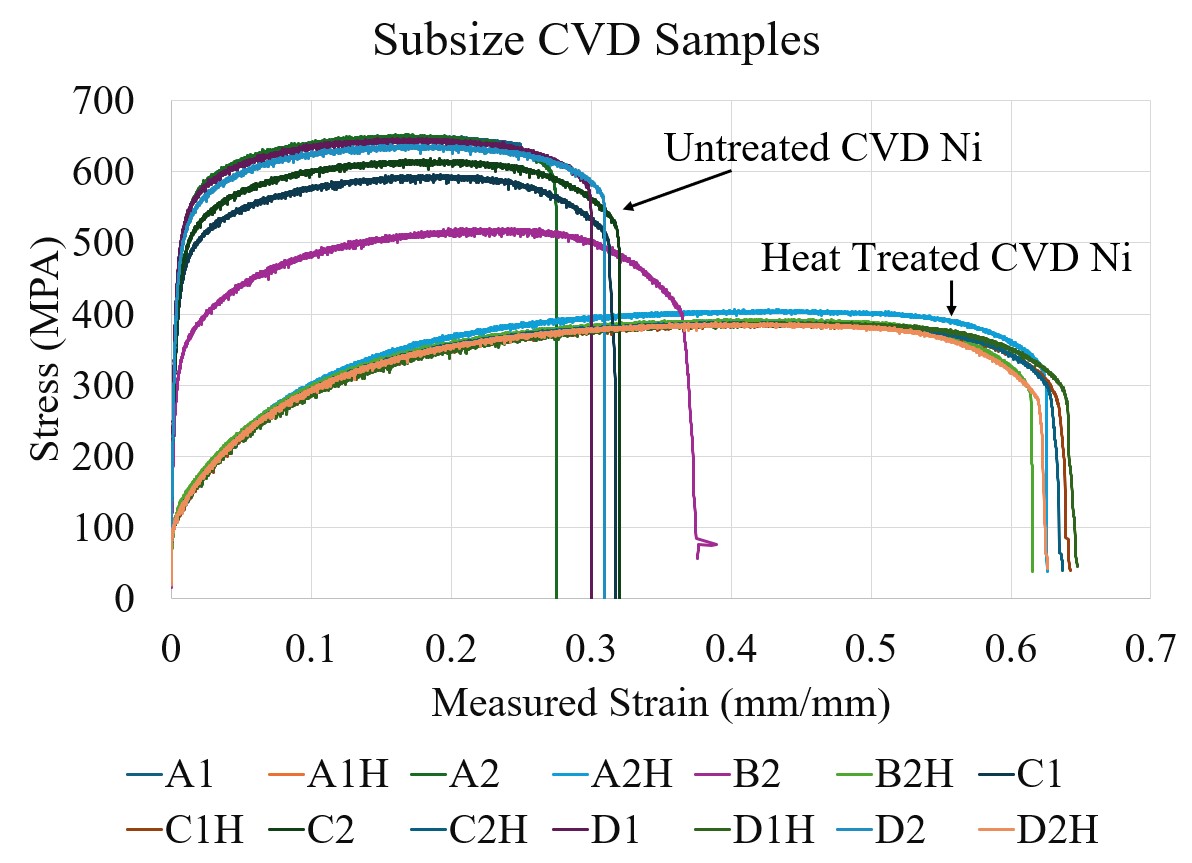}
    \caption{Tensile test curves for heat-treated CVD and COTS Ni samples, using the E8 subsize dog bones (See Fig.~\ref{fig:DogBones_small}).}
    \label{fig:small_pull}
\end{figure}

\subsection{Comparison to SNO NCD CVD Ni}
In~\cite{SNO-NCD-NIMA-2007} the SNO Collaboration reports several physical properties of CVD Ni including an ultimate tensile strength of $\sim$640~MPa. This is consistent with the results reported here for un-welded CVD Ni, and thus suggests consistency of the CVD Ni product. However, SNO Collaboration reports did not discuss weld-related tensile strength weakening or porosity at weld locations. This is notable given CVD Ni sections of the SNO NCD tubes were joined via laser welding with ``a 1054-nm Nd-YAG laser operating from 0.6 to 1.2 watts''~\cite{SNO-NCD-IEEE-1999}. The SNO Collaboration's positive welding results demonstrate mechan\-ically-sound welded structures of CVD Ni are feasible.

\section{Material Assay}\label{Sec:Assay}

In contemporary, low-radioactive background instrumentation, it is typical to seek materials containing parts per trillion (ppt)---\emph{or lower}---concentrations of contamination from naturally occurring radioactive species.\footnote{For reference 1~ppt of $^{232}$Th, $^{238}$U, and $^{\text{nat}}$K is equivalent to activity levels of 4.1, 12.4, and 31.6~$\mu$Bq/kg, respectively.} To systematically assess the CVD Ni for trace radioactive impurities, acid dissolution of the material followed by inductively coupled plasma mass spectrometry (ICP-MS) analysis was employed, as described below. COTS Ni was not evaluated for trace-level radioactive contamination in this study. A review of \url{www.radiopurity.org}~\cite{Loach_2016} reveals only five reports evaluating ``nickel'', all showing contamination levels orders of magnitude higher than seen in the bulk of the CVD Ni assessed in this report; this may warrant further investigation alongside future CVD Ni development.

\subsection{Analytical Methods}
Sub-samples of CVD Ni were processed to determine the trace-level contamination of $^{232}$Th, $^{238}$U, and $^{nat}$K. Analytical work was performed at PNNL in a Class 10000 cleanroom, and sample preparation was completed in a laminar flow hood providing a Class 10 environment. Optima grade nitric acid (HNO$_{3}$, Fisher Scientific, Pittsburg, PA, USA), Micro-90 \textsuperscript{\textregistered} detergent (Cole-Parmer, Vernon Hills, IL), and 100\% pure ethanol were used for all sample cleaning and preparation. Ultra-pure, 18.2~M$\Omega\cdot$cm water from a MilliQ system (Merk Millipore GmbH, Burlington, MA) was used for sample rinsing and in the preparation of reagent solutions. Ultralow background perfluoroalkoxy alkane (PFA) screw cap vials from Savillex (Eden Prairie, MN) were used as sample containers and as ICP-MS autosampler vials. All labware involved in sample handling and analysis (vials, tongs, pipette tips) were cleaned with 2\%~$v/v$ Micro-90 \textsuperscript{\textregistered} detergent, triply rinsed with MilliQ water, and leached in Optima grade 3M HCl and 6M HNO$_3$ solutions. Following leaching, all labware underwent validation to ensure cleanliness. The validation step consisted of pipetting 1.5~mL of 5\%~$v/v$ HNO$_3$ into each PFA vial. Vials were closed, shaken, and kept at 80~$\degree$C for at least 12~hours. Tongs and pipette tips were soaked in a 5\%~$v/v$ HNO$_3$ leaching solution (1.5~mL) for 5--10~minutes. The leachate from all labware was then analyzed via ICP-MS. The validation was performed to assure sufficiently low background for $^{232}$Th, $^{238}$U, and $^{nat}$K. Only labware for which signals were at reagent background levels passed validation. Labware failing validation underwent additional cycles of leaching and validation until background requirements were met. 

\subsection{ICP-MS Analysis} 
Determinations of Th, U, and K were performed using an Agilent 8900 ICP-MS (Agilent Technologies, Santa Clara, CA), equipped with an integrated autosampler, a microflow PFA nebulizer and a quartz double pass spray chamber. Plasma, ion optics, and mass analyzer parameters were tuned based on the instrumental response of a tuning standard solution from Agilent Technologies containing 0.1~ng/g Li, Co, Y, Ce, and Tl. In order to maximize signal-to-noise in the high $m/z$ range for Th and U analysis, the instrumental response for Tl ($m/z$ = 205) was used as a reference signal for instrumental parameter optimization. That is, the vendor-supplied tuning solution includes Tl ($m/z$ = 205), which has a $m/z$ and first ionization potential closest to those of Th and U. Oxides were monitored and kept below $2\%$ based on the $m/z = 156$ and $m/z = 140$ ratio from Ce (CeO$^{+}$/Ce$^{+}$) in the tuning standard solution. An acquisition method of three replicate samples and ten sweeps per replicate was used for each reading. Acquisition times for monitoring $m/z$ of interest (\textit{e.g.,} tracer and analyte isotopes) were set based on expected signals, in order to maximize instrumental precision by improving counting statistics.

Quantitation of $^{232}$Th and $^{238}$U was performed by isotope dilution methods, using the equation:
\begin{equation}
    \text{Concentration} = \dfrac{A_\text{analyte} \cdot C_\text{tracer}}{A_\text{tracer}} 
\end{equation}
where $A_\text{analyte}$ is the instrument response for the analyte, $A_\text{tracer}$ is the instrument response for the tracer and $C_\text{tracer}$ is the concentration of the tracer in the sample. Isotope dilution is the most precise and accurate method for quantitation for ICP-MS analysis, allowing the verification of sample preparation efficiency, and accounting for analyte losses and/or matrix effects. In this study non-naturally occurring $^{229}$Th and $^{233}$U were used as the tracer isotopes for the isotope dilution quantitation of $^{232}$Th and $^{238}$U, respectively. Absolute detection limits on the order of a few femtograms were attained for both $^{232}$Th and $^{238}$U. For bulk assay analysis, detection limits normalized to sample mass were on the order of 10--100~fg/g for both Th and U, while for surface contamination analysis, typically $< 1$~pg/g, corresponding to $< \mu$Bq/kg in terms of radioactivity for both bulk and surface contamination studies described below.

In order to achieve ultra-trace sensitivities for $^{232}$Th and $^{238}$U ion exchange chromatography was employed. This method improves analytical detection limits by preconcentrating the analytes of interest, in this case $^{232}$Th and $^{238}$U, and removing the majority of matrix components (\textit{i.e.}, Ni). In this work, the ion exchange procedure previously developed for ultra-sensitive assay of electroformed Cu was employed ~\cite{LAFERRIERE201593, ARNQUIST2020163761}. 

Determinations of natural potassium contamination levels were performed in cool plasma with NH$_3$ reaction mode. Instrumental parameters were optimized based on the instrumental response from a solution containing $\sim$1~ng/g $^\text{nat}$K with natural isotopic composition, diluted in-house from standard solutions. Quantitations of potassium were performed using an external calibration curve, generated using in-house diluted standard solutions with natural isotopic composition. 

Uncertainties for individual replicates were determined from the propagated uncertainties of the instrumental precision. Samples for which the analyte concentration was measured below the detection limit are reported as an upper limit. Values are reported in pg or fg per gram of sample for $^{232}$Th and $^{238}$U, and ng per gram for $^{nat}$K. The specific methods and processes described below are consistent with methods typically used for other matrix materials~\cite{LAFERRIERE201593,arnquist2023ultra}.

\subsection{Bulk Trace Contamination}

Initial assay efforts focused on evaluating trace radioactive contamination of the bulk interior of CVD Ni. To accomplish this, six replicates of $\sim$1~g sub-samples were selected for assay. Three replicates were taken from locations across each transect cut from the two CVD plates (see Fig.~\ref{fig:sampling_layout}). This approach was taken in order to sample diverse locations across the CVD material as a means of testing for heterogeneity across the CVD plates. Initial sample masses of the six replicates ranged from 1.23--1.44~g. To assess bulk trace radioactive contamination, each replicate of CVD Ni was etched (33--47\% by mass) using 4M HNO$_3$ to remove any surface contamination associated with handing and/or machining. The remaining mass of bulk CVD Ni was then digested stoichiometrically in 4M HNO$_3$ to produce concentrated Ni solutions in preparation for ion exchange chromatography. At this stage all sample preparation was conducted in tandem with process blanks in order to account for any contamination introduced during sample processing. Once fully dissolved, an aliquot of each replicate was diluted to $\sim$10000~ppm Ni using 2\% nitric in preparation for $^{\text{nat}}$K analysis. A matrix matched calibration curve was produced for K quantitation. During analysis of $^{\text{nat}}$K, samples were further diluted inline on the Agilent 8900 ICP-MS by a factor of $\sim$10. All remaining solutions were brought up to 8M HNO$_3$ using concentrated HNO$_3$ in preparation for ion exchange chromatography.

The measured concentrations of $^{232}$Th, $^{238}$U, and $^{\text{nat}}$K as well as their inferred activities are presented in Table~\ref{tab:BulkAssay}. Measured $^{232}$Th values for all but one replicate are above detection limit, with values ranging from 23--90~fg/g (0.09--0.37~$\mu$Bq/kg). Measured $^{238}$U values for all six bulk CVD Ni replicates are below the detection limit, $<$64--98~fg/g ($<$0.79--1.22~$\mu$Bq/kg). Measured potassium ($^{\text{nat}}$K) concentrations are all above detection limit with values ranging from 800--1000~pg/g (25--31~$\mu$Bq/kg of $^{40}$K). These bulk analysis results indicate that bulk CVD Ni contains very low levels of all three analyzed radio-contaminants, with the dominant source of radioactivity derived from $^{40}$K.

\renewcommand{\arraystretch}{1.2}
\begin{table*}
\centering
\begin{tabular}{| c | c | c | c | c c | c c | c c |} 
\hline 
 Ni transect & Initial mass & Bulk mass & Etch fraction & \multicolumn{2}{|c|}{$^{232}$Th} & \multicolumn{2}{|c|}{$^{238}$U} & \multicolumn{2}{|c|}{$^{\text{nat}}$K} \\
 & [g] & [g] & & [fg/g] & $\sigma$ & [fg/g] & $\sigma$ & [pg/g] & $\sigma$ \\
\hline
A & 1.2762 & 0.8270 & 35\% & $<$19 & $-$ & $<$98 & $-$ & 820 & 70 \\ 
\hline
A & 1.2363 & 0.7930 & 36\% & 23 & 6 & $<$71 & $-$ & 980 & 70 \\ 
\hline
A & 1.3017 & 0.8673 & 33\% & 90 & 6 & $<$64 & $-$ & 1000 & 100 \\ 
\hline
B & 1.2890 & 0.8540 & 34\% & 88 & 7 & $<$65 & $-$ & 900 & 200 \\ 
\hline
B & 1.3252 & 0.8250 & 38\% & 48 & 4 & $<$68 & $-$ & 850 & 100 \\ 
\hline
B & 1.4380 & 0.7659 & 47\% & 74 & 7 & $<$73 & $-$ & 800 & 100 \\
\hline 
\end{tabular}

\caption{Contamination levels of $^{232}$Th, $^{232}$U, and $^{\text{nat}}$K in CVD Ni samples. The Ni transects A and B refer to the two CVD Ni plates (see Sec.~\ref{Sec:Samples} and Fig.~\ref{fig:sampling_layout}). Concentration equivalent units: [fg/g]$=$ppq and [pg/g]$=$ppt.}
\label{tab:BulkAssay}
\end{table*}
\renewcommand{\arraystretch}{1.0}

\subsection{Surface Trace Contamination}
A key concern in use of manufacturer-supplied or highly processed structural materials is the potential for processing and handling steps to introduce additional trace radioactive species onto or into the surface of materials. For CVD Ni, a past study identified the substrate on which the Ni was deposited as a potential vector for contamination on the surface of the Ni (see Sec.~\ref{sec:SNO-CVD-assays} for further details). To investigate this possibility, acid etching was used to sample the surface matrix of the CVD Ni. Two studies were performed on two sets of three sub-samples to understand the surface concentrations and depth profile of $^{232}$Th, $^{238}$U, and $^{40}$K: sequential etches of $\sim$1\% by mass each (four etches total), and stoichiometric etches of $\sim1$~$\mu$m each (10 etches total) for better granularity of the outer 1\%. The sub-samples from the second study were then subjected to three more etches (two 1\% etches followed by 25\%, all by mass), followed by dissolution, ion chromatography, and analysis to compare surface cleaned sample contamination levels with bulk assay study results. Both studies were conducted in tandem with three process blanks.

For the first study, three 0.5~cm$^{3}$ ($\sim$1~g each) CVD Ni sub-samples were randomly selected from the CVMR Ni panels (Figs.~\ref{fig:as_received} and~\ref{fig:sampling_layout}). Sub-samples were initially cleaned by sonication for 15~minutes in 2\% $v/v$ Micro-90 detergent followed by ultrapure ethanol (triply rinsed in ultrapure water after each) for removal of contamination due to handling (\textit{e.g.}, fingerprints and/or dust). Each sub-sample was then etched in 2~mL 4M HNO$_{3}$ for 4--21~hours with the goal of removing $\sim$1\% by mass with each etch. After each etch, samples were triply rinsed with ultrapure water and dried with ultra high purity (UHP) N$_{2}$ gas. Four etches were conducted in the same fashion, with a total of $\sim$10\% mass etched from each sub-sample. Etchant solutions were diluted to $\sim$700~ppm Ni for ICP-MS analysis.

The second study was conducted using the same meth\-od\-ology as the first, but with three new sub-samples and 0.0152~mL of 4 M HNO$_{3}$---the exact amount required to stoichiometrically etch 1~$\mu$m from each surface of each cube, assuming uniform mass removal during etching. Ten sequential etches were performed to look more closely at the outer $\sim$1\%, where the majority of the contamination was observed in the first study (see Figure~\ref{fig:etch}). As noted above, CVD Ni from CVMR is deposited onto an Al substrate and a prior study showed that Al contaminates the substrate-side of the CVD Ni surface~\cite{SNO-STR-2000-005}. To understand whether there is a correlation between the Al surface contamination (as previously observed by the SNO Collaboration), and surface contamination from Th, U, and K presented in this report, the presence of trace-levels of Al in the CVD Ni was also measured in the samples studied here. Thus, while Th, U, and K contamination is the primary focus of this report, the Al serves as both a likely tracer for these contaminants, as well as potentially creating an additional concern for specific high-purity applications.

The results of the two surface etching procedures are presented in Figure~\ref{fig:etch}. Figure~\ref{fig:aluminum} presents the results for Al in the second, stoichiometric etching tests. These results show profiles of contamination on the surface and near-surface depths. The highest concentrations of impurities are seen nearest to the surface, as evidenced by the first of the stoichiometric etches which aimed to remove the outermost $\sim$1~$\mu$m of material for analysis. This is seen in panels C, I, and O in Figure~\ref{fig:etch} and the left most three data points in Figure~\ref{fig:aluminum}. For comparison, the anticipated inner concentration levels from the analysis reported in Table~\ref{tab:BulkAssay} are shown as the ``Bulk''-labeled gray bands in Figure~\ref{fig:etch}.

These results strongly suggest surface-related Th, U, and K contamination extends at least 10~$\mu$m into the CVD Ni and likely deeper. This latter conjecture stems from panels F, L, and R in Figure~\ref{fig:etch} showing that even after $\sim$28\% surface removal, the Th, U, and K contamination remains above the bulk concentration levels. If it is hypothesized that this near-surface contamination is driven by the Al growth substrate, then it is important to recognize (1) it is likely only the substrate side of the material that actually contains elevated contamination levels and (2) the etching process \emph{did not} target a single side of the material. As the work presented in this report did not selectively target the Al substrate side of the CVD Ni samples, it is recommended that future impurity contamination measurements seek the ability to provide that discrimination capability. For the present work, these are important caveats for assessing the absolute concentration values reported at any depth within the CVD Ni.

Another significant caveat arises: The sequential etching approach employed assumes uniform mass removal from all acid-exposed surfaces. In fact, during the second stoichiometric etching tests, it was observed that material removal during the incremental etches did not appear uniform. More specifically, there was visual observation of heterogeneous removal of the external oxidation layer (see Fig.~\ref{fig:etchphoto}).

Despite the identified caveats, the preponderance of evidence points to surface and near-surface contamination in the CVD Ni. Further investigation is warranted to determine the precise distribution of surface-related radio-contaminants. More controlled methods (\textit{e.g.}, initial plasma etching, electrochemistry, or thermal reduction of outer oxide layer followed by subsequent acid etching) would greatly enhance the precision determination of contamination as a function of position within the material.

\begin{figure}
    \centering
    \includegraphics[width=0.8\linewidth]{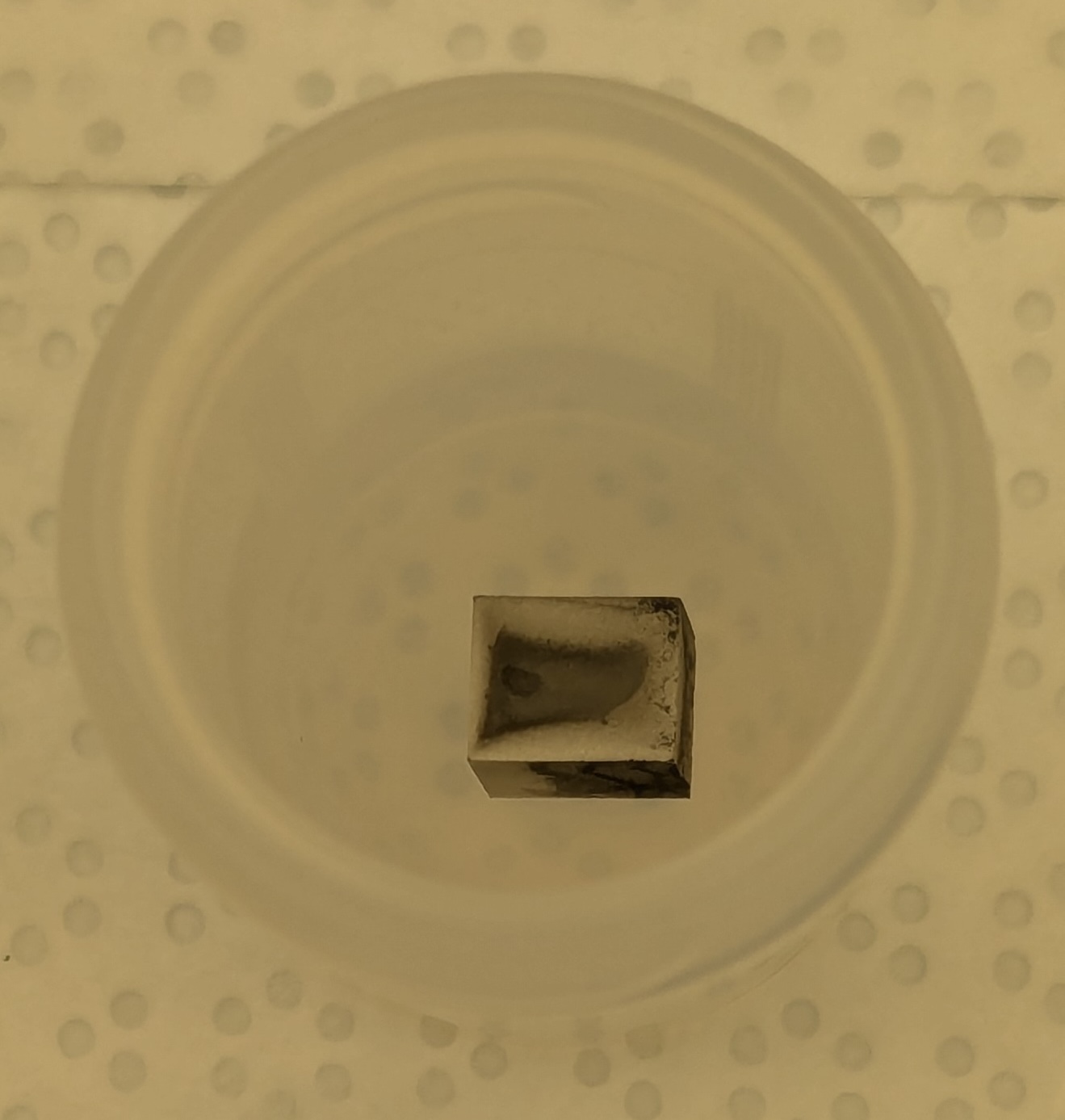}
    \caption{Sub-sample B-3A1 after etch 9. Note outer, darker oxidation layer is still present in places, but not in others. This could represent non-uniform removal of material, or non-uniform oxidation of the surface.}
    \label{fig:etchphoto}
\end{figure}

\begin{figure*}[h!tbp]
    \centering
    \includegraphics[width=1\linewidth]{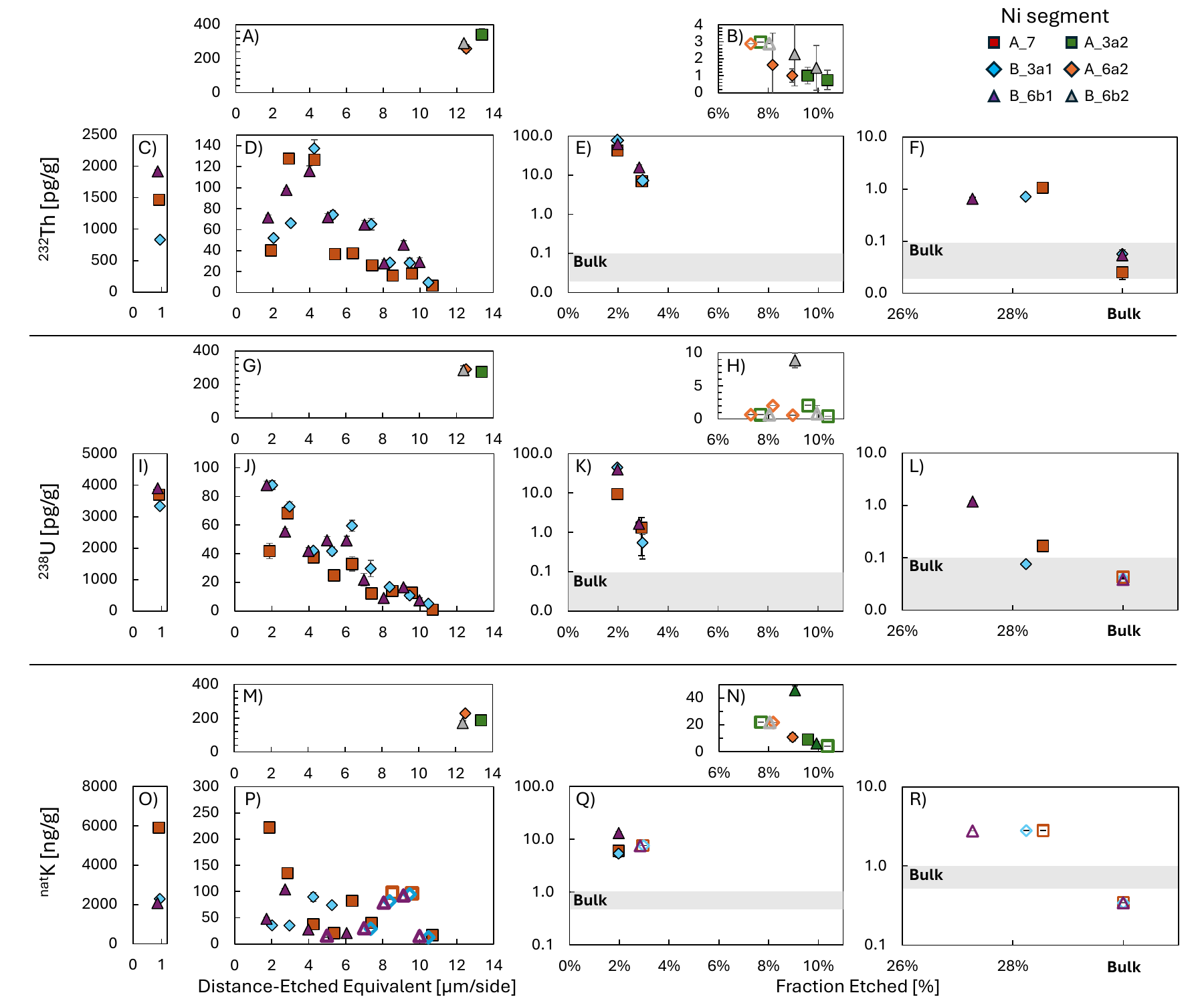}
    \caption{Results for both surface etching studies. The $^{232}$Th and $^{238}$U values are reported in pg/g (ppt) and $^{nat}$K is reported as ng/g (ppb). Panels A--B, G--H, and M--N show results from the initial coarse etching study, while all other panels show data from the second stoichiometric etching study. The shaded gray regions labeled ``Bulk'' reflect the range of measured bulk radioactivity values reported in Table~\ref{tab:BulkAssay}. The measured contamination concentrations for the outer 15~$\mu$m are presented as an equivalent distance etched [$\mu$m/side] for each etch aliquot, calculated from total mass removed for a given etch and \emph{assuming} uniform mass removal. For deeper etch results, the etch is presented as the mass fraction etched expressed as a percentage of the initial Ni sub-sample mass. Open symbols denote upper limits and error bars represent instrument uncertainty (1$\sigma$).}
    \label{fig:etch}
\end{figure*}

\begin{figure}
    \centering
    \includegraphics[width=1\linewidth]{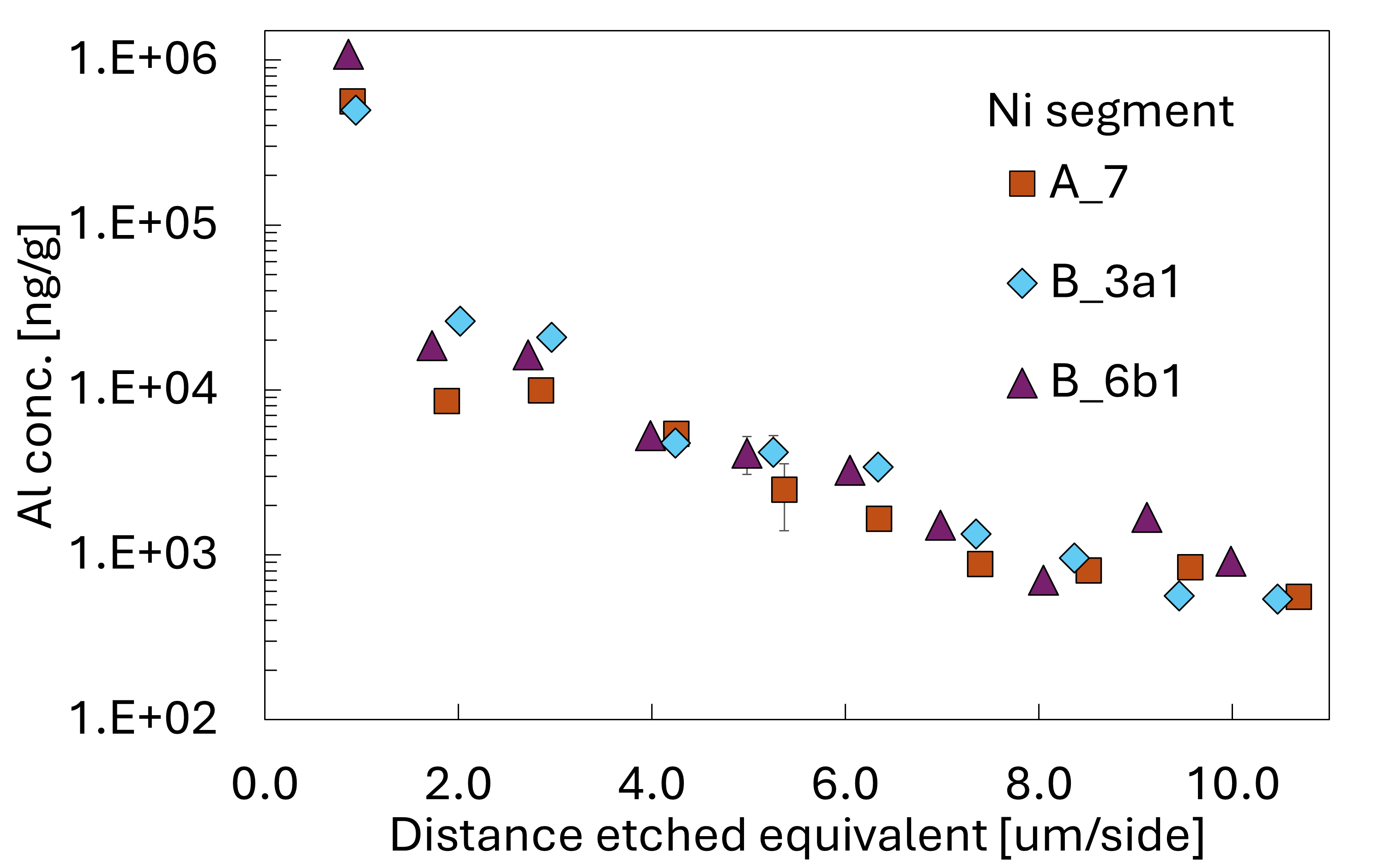}
    \caption{Results for Al concentration in ng/g (ppb) for sub-samples etched in second study.}
    \label{fig:aluminum}
\end{figure}

\subsection{Comparison to SNO NCD CVD Ni}\label{sec:SNO-CVD-assays}
The SNO Collaboration's reports on the trace radioactive contamination within the \emph{bulk} of CVD Ni e\-volv\-ed over time. An initial 1993 report~\cite{SNO-STR-93-054} using radiochemical neutron-activation analysis (RNAA) reported concentrations of Th at 7.4$\pm$0.4~ppt and U at $\leq$0.8~ppt~\footnote{The authors are assuming the ``M-Tube'' sample reported in~\cite{SNO-STR-93-054} is representative of the Mirotech material chosen for the SNO NCDs.}. A pair of later theses~\cite{Browne1999,Stonehill2005} used methods of proportional counter ``self-counting'', combined with radiation transport modeling, to assess the trace Th and U contamination in the NCD CVD Ni. The theses arrived at best estimate values of less than 10~ppt for both Th and U; values which were typically higher than, though strictly speaking consistent with, the RNAA results. A 2007 summary design publication~\cite{SNO-NCD-NIMA-2007} presents $1 \pm 1$~pg/g of $^{232}$Th and $<5$~pg/g of $^{238}$U in the NCD CVD Ni bodies, stated as determined through a combination of RNAA and high purity germanium (HPGe) gamma-ray spectroscopy counting. The publication also mentions the higher values Th and U concentrations attributed from the ``self-counting'' methods. Both theses predate the design publication and a recognition there was additional \emph{surface} contamination ``hot spots'' on the outer surface of an NCD, see~\cite{Wan2008,OKeeffe2008,SNO-NCD-NIMA-2011}. A later thesis~\cite{OKeeffe2008} states the recommended values for the bulk contamination of the CVD Ni in the NCDs is $3.43^{+1.49}_{-2.11}$~pg/g of $^{232}$Th and $1.81^{+0.80}_{-1.12}$~pg/g of $^{238}$U based upon a maximum likelihood analysis of four dominant backgrounds measured within the SNO detector heavy water volume.

The results presented in this report consistently show tens of femtogram/gram (fg/g) contamination of $^{232}$Th, and less than those levels for $^{238}$U. These determinations are more than an order of magnitude lower than reported by the SNO Collaboration. There are several plausible hypotheses for this significant discrepancy: (1) After $\sim$30~years the CVD Ni production process has changed, (2) The difference in the surface-to-volume ratio of analyzed samples is an important factor (see surface contamination discussion below), and (3) The SNO Collaboration was not able to fully disentangle the bulk and surface contamination contributions through available methods. These hypotheses are not mutually exclusive and all may contribute.

The examination of the surface-depth profile of contamination presented in this report apparently confirms the SNO Collaboration's determination that Al contamination of the substrate-side of the CVD Ni surface carries trace Th and U contamination (see Figs.~\ref{fig:etch} and~\ref{fig:aluminum}). The results presented above in this report are consistent with prior analysis of the profile of Al in the interior surface of the SNO NCD CVD Ni tubes~\cite{SNO-STR-2000-005}; though the absolute Al concentration report is slightly lower than that reported by the SNO Collaboration. This may be a result from a difference in the sample geometry used in the progressive surface etching employed in both cases. That is, the SNO NCD tubes provided a two-sided, ``hoop'' geometry (nominal radial CVD Ni thickness of 370~$\mu$m), whereas the sampling used in this report was a six-sided, ``cube'' geometry (nominally 0.5~cm-sided cubes). This sample geometry difference could account for a factor of $\sim$3 lower concentration when comparing the Al concentration results presented in this report to the SNO NCD Al concentration results. Additionally, results in this report showed that surface contamination from Th, U, K, and Al extends into the CVD Ni surface, perhaps more than 10~$\mu$m, \emph{``as supplied'' by the manufacturer}. As noted in this report's Introduction (Sec.~\ref{sec:Intro}), the SNO Collaboration used electropolishing and etching as a surface cleaning step to remove material, which occurred \emph{prior} to the SNO NCD CVD Ni contamination analyzes for Th and U. Thus, while the SNO Collaboration's reports and this report are consistent regarding the existence of surface and substrate-related contamination of the CVD Ni, there are enough differences in the details of the measurement processes to make positing an explanatory relationship between the quantitative results very difficult. 

A few additional notes on the SNO NCDs are provided for completeness. During the NCD fabrication phase the CVD Ni was stored in an underground location without adequate protection from plate-out of radon progeny. As a result, efforts were taken to clean the surfaces of the NCDs prior to deployment. Additionally, the SNO NCDs spent two or more years deep underground prior to deployment in a ``cool-down'' phase to allow decay of $^{56}$Co (77~day half-life) cosmogenically produced in the CVD Ni bulk during surface-based fabrication. While these details are in general important for using CVD Ni as a low radioactive background material, neither of these SNO NCD radiopurity issues are directly related to the material assay results presented in this report.

\section{Conclusion}\label{Sec:Conclusion}

The strength of a CVD Ni structure will likely be determined by the quality of the weld joints. The SNO Collaboration's laser welding technique proved effective, though the authors of this report did not find any information on the ultimate tensile strength of the SNO NCD's welded CVD Ni joints. Based on the investigations presented in this report, it is anticipated that joints of welded CVD Ni will be at most as strong as the ultimate tensile strength seen in COTS Ni. Further development of the welding process for CVD Ni is warranted to determine an ideal weld process for large mechanical structures that can demonstrate consistent ultimate tensile strength through repeated evaluation.

Furthermore, future work is recommended to assess the impact of CVD Ni welding processes on the trace levels of radioactive species present at the weld joint. To achieve this, referring to the sample flow diagram Figure~\ref{fig:flow_diagram}, some samples processed through the ``Welding'' process box should move to the ``Acid Dissolution'' process box for measurement of radioactive species in the ``ICP-MS Assay'' process box.

Based on the impurity depth profile seen in Figure~\ref{fig:etch}, it is recommended that future CVD Ni development consider using a higher purity substrate for Ni deposition. There are a few suggested approaches: (1) Use a higher purity Al substrate, (2) Use a material substrate other than Al (which is higher purity), and (3) In all cases, analyze the substrate material for Th, U, and K in advance of CVD Ni production.

Furthermore, the assay measurements presented in this report \emph{only} assessed the \emph{tops} of the $^{232}$Th and $^{238}$U decay chains. It is often the case that decays from intermediate progeny in these decay series generate the primary background concern for a given experiment, depending on the details of experimental design. It is very likely that the CVD Ni manufacturing process disturbs the equilibrium of the decay chains. Thus, it is recommended that future material assay assessments of CVD Ni target the specific isotopes of concern for a given experimental design, in addition to the $^{232}$Th, $^{238}$U, and $^{\text{nat}}$K studied in this work.

In summary, this report presents an initial evaluation of CVD Ni for its utility in the construction of large mechanical structures of low-radioactive background experiments. The material appears promising from both strength and low levels of rad\-io\-ac\-tive con\-tam\-i\-nants, though further development is warranted for any specific application.

\section*{Acknowledgments}
The authors are grateful to CVMR for collaborating to produce the materials used in this study. The authors appreciate added insight on the SNO NCD CVD Ni history provided by P.J.~Doe, N.S.~Oblath, and R.G.H.~Ro\-bert\-son. The authors thank A.~Gilbert for performing x-ray the examination.

This work was supported in part by Laboratory Directed Research and Development (LDRD) programs at Brookhaven National Laboratory (BNL), Lawrence Livermore National Laboratory (LLNL), Pacific Northwest National Laboratory (PNNL), and SLAC National Accelerator Laboratory. The authors gratefully acknowledge support for nEXO from the Office of Nuclear Phys\-ics within DOE’s Office of Science under grants/con\-tracts DE-\-AC02-\-76SF00\-515, DE-\-FG02-\-01ER\-41\-166, DE-\-SC\-00\-2305, DE-\-FG02-\-93ER\-40\-789, DE-\-SC\-00\-21\-388, DE-\-SC\-00\-12\-704, DE-\-AC52-\-07NA\-27\-344, DE-\-SC\-00\-17\-970, DE-\-AC05-\-76RL\-01\-830, DE-\-SC\-00\-12\-654, DE-\-SC\-00\-21\-383, DE-\-SC\-00\-14\-517, DE-\-SC\-00\-24\-666, DE-\-SC\-00\-20\-509, DE-\-SC\-00\-24\-677 and support by the US National Science Foundation grants NSF PHY-\-2111\-213 and NSF PHY-\-2011\-948; from NSERC SAPPJ-\-2022-\-00021, CFI 39881, FR\-QNT 2019-\-NC-\-255821, and the CFREF Ar\-thur B. McDonald Canadian Astroparticle Physics Research Institute in Canada; from IBS-\-R016-\-D1 in South Korea; and from CAS in China.

%
%



\bibliographystyle{elsarticle-num-jlo} 
\bibliography{mainbibfile} 

\begin{thebibliography}{10}
\expandafter\ifx\csname url\endcsname\relax
  \def\url#1{\texttt{#1}}\fi
\expandafter\ifx\csname urlprefix\endcsname\relax\def\urlprefix{URL }\fi
\expandafter\ifx\csname href\endcsname\relax
  \def\href#1#2{#2} \def\path#1{#1}\fi

\bibitem{nEXO_Adhikari_2022}
G~Adhikari, et~al., {nEXO}: Neutrinoless double beta decay search beyond
  10$^{28}$ year half-life sensitivity, Journal of Physics G: Nuclear and
  Particle Physics 49~(1) (2021) 015104.
\newblock \href {https://doi.org/10.1088/1361-6471/ac3631}
  {\path{doi:10.1088/1361-6471/ac3631}}.

\bibitem{Majorana2006}
Ettore Majorana and Luciano Maiani, A symmetric theory of electrons and
  positrons, Springer Berlin Heidelberg, Berlin, Heidelberg, 2006, pp.
  201--233.
\newblock \href {https://doi.org/10.1007/978-3-540-48095-2_10}
  {\path{doi:10.1007/978-3-540-48095-2_10}}.

\bibitem{WOS:000492825700009}
Michelle~J. Dolinski, Alan W.~P. Poon, and Werner Rodejohann, Neutrinoless
  double-beta decay: Status and prospects, in: BR~Holstein (Ed.), ANNUAL REVIEW
  OF NUCLEAR AND PARTICLE SCIENCE, VOL 69, Vol.~69 of Annual Review of Nuclear
  and Particle Science, 2019, pp. 219--251.
\newblock \href {https://doi.org/10.1146/annurev-nucl-101918-023407}
  {\path{doi:10.1146/annurev-nucl-101918-023407}}.

\bibitem{nEXO-PreCDR}
nEXO Collaboration, et~al., {nEXO} pre-conceptual design report (2018).
\newblock \href {http://arxiv.org/abs/1805.11142} {\path{arXiv:1805.11142}}.

\bibitem{HALO_2008}
C~A Duba, et~al., {HALO} – the helium and lead observatory for supernova
  neutrinos, Journal of Physics: Conference Series 136~(4) (2008) 042077.
\newblock \href {https://doi.org/10.1088/1742-6596/136/4/042077}
  {\path{doi:10.1088/1742-6596/136/4/042077}}.

\bibitem{BOGER2000172}
J~Boger, et~al., {The Sudbury Neutrino Observatory}, Nuclear Instruments and
  Methods in Physics Research Section A: Accelerators, Spectrometers, Detectors
  and Associated Equipment 449~(1) (2000) 172--207.
\newblock \href {https://doi.org/https://doi.org/10.1016/S0168-9002(99)01469-2}
  {\path{doi:https://doi.org/10.1016/S0168-9002(99)01469-2}}.

\bibitem{PhysRevC.88.025501}
B.~Aharmim, et~al., Combined analysis of all three phases of solar neutrino
  data from the {Sudbury Neutrino Observatory}, Phys. Rev. C 88 (2013) 025501.
\newblock \href {https://doi.org/10.1103/PhysRevC.88.025501}
  {\path{doi:10.1103/PhysRevC.88.025501}}.

\bibitem{SNO-NCD-IEEE-1999}
MC~Browne, et~al., Low-background $^3$he proportional counters for use in the
  {Sudbury Neutrino Observatory}, IEEE TRANSACTIONS ON NUCLEAR SCIENCE 46~(4,
  1) (1999) 873--876, 1998 Nuclear Science Symposium (NSS), TORONTO, CANADA,
  NOV 08-14, 1998.
\newblock \href {https://doi.org/10.1109/23.790695}
  {\path{doi:10.1109/23.790695}}.

\bibitem{SNO-NCD-NIMA-2007}
J.~F. Amsbaugh, et~al., Array of low-background $^3$he proportional counters
  for the {Sudbury Neutrino Observatory}, NUCLEAR INSTRUMENTS \& METHODS IN
  PHYSICS RESEARCH SECTION A-ACCELERATORS SPECTROMETERS DETECTORS AND
  ASSOCIATED EQUIPMENT 579~(3) (2007) 1054--1080.
\newblock \href {https://doi.org/10.1016/j.nima.2007.05.321}
  {\path{doi:10.1016/j.nima.2007.05.321}}.

\bibitem{CVMR}
CVMR Corporation, 35 Kenhar Drive, Toronto, ON, Canada. M9L 1M9.
  \url{https://cvmr.ca/}.

\bibitem{MondCVD-1890}
Ludwig Mond, Carl Langer, and Friedrich Quincke, L.—action of carbon monoxide
  on nickel, J. Chem. Soc.{,} Trans. 57 (1890) 749--753.
\newblock \href {https://doi.org/10.1039/CT8905700749}
  {\path{doi:10.1039/CT8905700749}}.

\bibitem{Bansa2001}
Patrice~B. Bansa, Property characterization of {CVD} nickel, Master's thesis,
  University of Toronto (2001).

\bibitem{Loach_2016}
J.C. Loach, et~al., A database for storing the results of material radiopurity
  measurements, Nuclear Instruments and Methods in Physics Research Section A:
  Accelerators, Spectrometers, Detectors and Associated Equipment 839 (2016)
  6–11.
\newblock \href {https://doi.org/10.1016/j.nima.2016.09.036}
  {\path{doi:10.1016/j.nima.2016.09.036}}.

\bibitem{LAFERRIERE201593}
B.D. LaFerriere, T.C. Maiti, I.J. Arnquist, and E.W. Hoppe, A novel assay
  method for the trace determination of th and u in copper and lead using
  inductively coupled plasma mass spectrometry, Nuclear Instruments and Methods
  in Physics Research Section A: Accelerators, Spectrometers, Detectors and
  Associated Equipment 775 (2015) 93--98.
\newblock \href {https://doi.org/https://doi.org/10.1016/j.nima.2014.11.052}
  {\path{doi:https://doi.org/10.1016/j.nima.2014.11.052}}.

\bibitem{ARNQUIST2020163761}
I.J. Arnquist, M.L. {di Vacri}, and E.W. Hoppe, An automated ultracleanion
  exchange separation method for the determinations of 232th and 238u in copper
  using inductively coupled plasma mass spectrometry, Nuclear Instruments and
  Methods in Physics Research Section A: Accelerators, Spectrometers, Detectors
  and Associated Equipment 965 (2020) 163761.
\newblock \href {https://doi.org/https://doi.org/10.1016/j.nima.2020.163761}
  {\path{doi:https://doi.org/10.1016/j.nima.2020.163761}}.

\bibitem{arnquist2023ultra}
Isaac~J Arnquist, et~al., Ultra-low radioactivity flexible printed cables, EPJ
  Techniques and Instrumentation 10~(1) (2023) 17.

\bibitem{SNO-STR-2000-005}
Miles Smith, Aluminum contamination of nickel tubes used in {NCD} construction,
  SNO Technical Report 005, University of Washington (2000).

\bibitem{SNO-STR-93-054}
G.~G. Miller, J.~A. Olivares, and J.~B. Wilhelmy, Determination of {Th} and {U}
  in high purity {Ni}, SNO Technical Report 054, Los Alamos National Laboratory
  (1993).

\bibitem{Browne1999}
Michael~Charles Browne, Preparation for deployment of the {Neutral Current
  Detectors (NCDs)} for the {Sudbury Neutrino Observatory}, Ph.D. thesis, North
  Carolina State Washington (1999).

\bibitem{Stonehill2005}
Laura~C. Stonehill, Deployment and background characterization of the {Sudbury
  Neutrino Observatory Neutral Current Detectors}, Ph.D. thesis, University of
  Washington (2005).

\bibitem{Wan2008}
Hok Seum Wan~Chan Tseung, Simulation of the {Sudbury Neutrino Observatory
  Neutral Current Detectors}, Ph.D. thesis, Wadham College, Oxford (2008).

\bibitem{OKeeffe2008}
Helen~Mary O'Keeffe, Low energy backgorund in the {NCD Phase} of the {Sudbury
  Neutrino Observatory Neutral Current Detectors}, Ph.D. thesis, Lincoln
  College, Oxford (2008).

\bibitem{SNO-NCD-NIMA-2011}
H.~M. O'Keeffe, et~al., Four methods for determining the composition of trace
  radioactive surface contamination of low-radioactivity metal, NUCLEAR
  INSTRUMENTS \& METHODS IN PHYSICS RESEARCH SECTION A-ACCELERATORS
  SPECTROMETERS DETECTORS AND ASSOCIATED EQUIPMENT 659~(1) (2011) 182--192.
\newblock \href {https://doi.org/10.1016/j.nima.2011.08.060}
  {\path{doi:10.1016/j.nima.2011.08.060}}.

\end{thebibliography}






\end{twocolumn} 

\end{document}